\documentclass[10pt,pre,aps,twocolumn,floatfix,superscriptaddress]{revtex4-1}

\usepackage[utf8]{inputenc}
\usepackage{amsmath}
\usepackage{graphicx,color}
\usepackage{mathptmx}

\usepackage{tikz}
\usepackage{environ}
\usepackage{layouts}

\usetikzlibrary{external,positioning} 
\begin{document}
\title{Ultra-Intense Laser Pulses in Near-Critical Underdense Plasmas --- Radiation Reaction and Energy Partitioning}

\author{Erik Wallin} 
\affiliation{Department of Physics, Chalmers University of Technology, SE--412 96 G\"oteborg, Sweden}

\author{Arkady Gonoskov} 
\affiliation{Department of Physics, Chalmers University of Technology, SE--412 96 G\"oteborg, Sweden}
\affiliation{Institute of Applied Physics, Russian Academy of Sciences, Nizhny Novgorod 603950, Russia}
\affiliation{University of Nizhny Novgorod, Nizhny Novgorod 603950, Russia}

\author{Christopher Harvey}
\affiliation{Department of Physics, Chalmers University of Technology, SE--412 96 G\"oteborg, Sweden}

\author{Olle Lundh}
\affiliation{Department of Physics, Lund Institute of Technology, P.O. Box 118,SE--221 00 Lund, Sweden}

\author{Mattias Marklund}
\affiliation{Department of Physics, Chalmers University of Technology, SE--412 96 G\"oteborg, Sweden}

\begin{abstract}
Although, for { \it current} laser pulse energies, the weakly nonlinear regime of LWFA is known to be the optimal for reaching the highest possible electron energies, the capabilities of upcoming large laser systems will provide the possibility of running highly nonlinear regimes of laser pulse propagation in underdense or near-critical plasmas.  Using an extended particle-in-cell (PIC) model that takes into account all the relevant physics, we show that such regimes can be implemented with external guiding for a relatively long distance of propagation and allow for the stable transformation of laser energy into other types of energy, including the kinetic energy of a large number of high energy electrons and their incoherent emission of photons. This is despite the fact that the high intensity of the laser pulse triggers a number of new mechanisms of energy depletion, which we investigate systematically.
\end{abstract}

\maketitle

\section{Introduction}
Laser wakefield acceleration (LWFA) is a means for utilising the extreme fields accessible in plasmas for the purpose of accelerating electrons to very high energies over short distances \cite{tajima-dawson}. Originally starting out as a basic principle of electron acceleration, this field has matured into one with a manifold of applications \cite{malka_rmp,malka2008principles}. As laser technology has developed, so have the attainable laser intensities and repetition rates. A natural question is therefore if the progress in the field of LWFA is likely to continue as we ramp up the power of the laser facilities, or if new physical phenomena may affect the development of laser-driven electron accelerators. Here we try to shed light on this question by invoking large scale particle-in-cell (PIC) simulations of LWFA. Indeed, we find that there are effects that will  alter the acceleration process as we increase the laser intensity of the pulses used to generate the wakefield. Moreover, due to the nature of these processes, we find that there can also be certain benefits, such as an efficient energy transformation from the optical range to the range of XUV and gamma rays. \cite{malka_rmp}. We therefore expect that, guided by our simulations and analytical understanding, new possibilities may open up for future LWFA systems. 

Recent, and planned, developments in laser systems \cite{vulcan, eli,xcels} are aimed at reaching new regimes for laser-matter interactions \cite{mourou-rmp}. The laser radiation of the expected intensities will trigger new semi-classical (or even quantum electrodynamical (QED)) effects\cite{marklund-shukla,Heinzl:2008an,dipiazza-rmp}, that should be taken into account in theoretical analysis and computational models. Furthermore, computational models are inherently limited by their finite resolution, posing limits to their validity. This can in principle lead to faulty estimates when considering certain physical effects.  One such effect is that of radiation reaction (RR) \cite{VranicRR, radiationreaction} in PIC simulations, where the problem is due to the limitations in resolving the emitted high frequency radiation \cite{wallin2015}. The correct inclusion of such an effect can change the output from e.g. a laser wakefield accelerator.

\begin{figure}
  \includegraphics[width=1.0\textwidth]{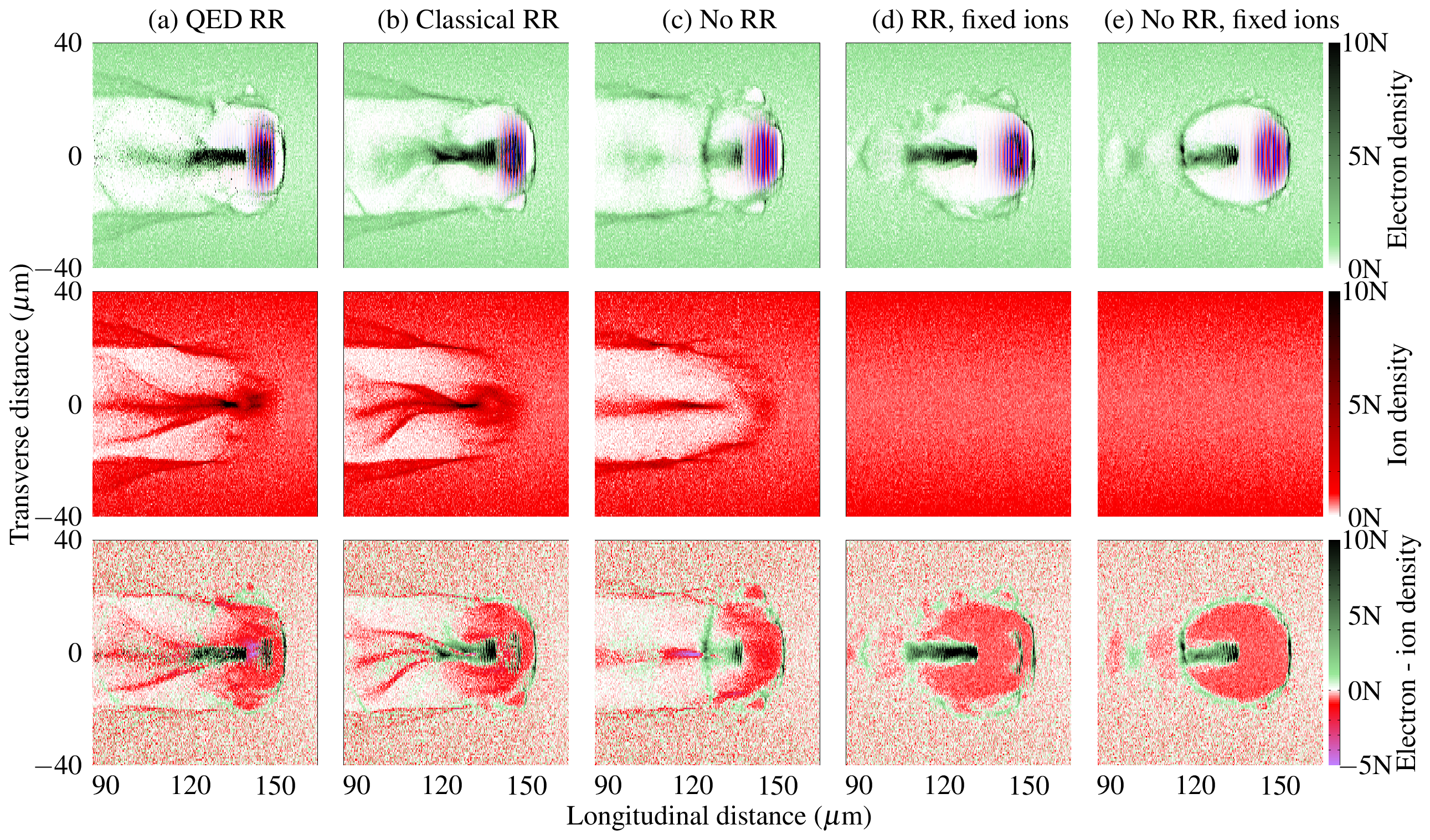}
  \caption{Comparing the effects of classical and quantum radiation reaction in a ultra-high intensity regime. Here $I = 4 \times 10^{24} \, \mathrm{W \, cm^{-2}}$ and $N = 3 \times 10^{20}\, \mathrm{cm}^{-3}$. In (a) we show the case of QED radiation reaction, in (b) classical radiation reaction via the Landau-Lishitz model, and in (c) the case without radiation reaction. All these three cases have self-consistent ion motion included. In (d) we have included classical radiation reaction via the Landau-Lifshitz model but kept the ions (hydrogen) stationary and in case (e) the ions are fixed and the radiation reaction is turned off. As expected, the notion of fixed ions gives results very far from the self-consistent picture. We note a significant effect of radiation reaction on the ion motion by comparison of (a), (b), and (c), while the difference in (a) and (b) is negligible. Thus, the small differences between (a) and (b) supports the notion that we in the LWFA case can use the classical radiation reaction model with great accuracy even at very high intensities, putting our use of the classical model for radiation reaction for these systems on a firm footing.}
\label{fig:Classical-QED-RR}
\end{figure}

In our simulations, we overcome this problem via the utilisation of (i) a radiation reaction module as well as (ii) a synchrotron module. We futhermore include the process of pair production via a Breit-Wheeler process. Certainly, at high enough intensities of the laser radiation, this classical description is not applicable as the energies of electrons and emitted photons become comparable. One then needs a quantum description of the emission process, which implies probabilistic generation of a photon accompanied with a respective recoil. The major differences between these descriptions are the stochastic properties of photon emission and the discrete nature of the photons, which can for example lead to a broadening of the spatial distribution of photon emission. By comparing our numerical results with the ones obtained using the further extended QED-PIC model \cite{QEDPIC} we identify the range of intensities $< 10^{25}$~W/cm${^2}$, where the outlined effects do not contribute substantially enough to affect our conclusions (see Fig.\ \ref{fig:Classical-QED-RR}). In particular, we do \emph{not} find any significant pair production in these systems for intensities up to $\sim 10^{26} \mathrm{W\, cm^{-2}}$ (this is to be expected, as there are very few head-on collisions between the laser field and the electrons when the electron distribution is close to thermal). Throughout the current study we consider this intensity range and therefore use PIC code extended with the inclusion of models (i) and (ii), leaving the related QED effects out of the scope of the present paper.

\subsection{Laser wakefield acceleration}
In laser wakefield acceleration \cite{tajima-dawson,Sprangle1988} a short, high intensity pulse propagates through an underdense plasma. Electrons are pushed to the sides of the laser pulse, with some of them ending up as a bunch of electrons in the wake behind the laser. The electron cavity induces a strong electric field (hundreds of GV/m \cite{Leemans2002,Malka2002, Modena1995}), accelerating the electrons behind the laser. In the bunch, the electrons undergo transverse oscillations \cite{Kiselev2004}, emitting high frequency radiation in the process. The output is a highly collimated, focused and quasi-monoenergetic electron beam \cite{Geddes2004, leemans2006gev}, as well as high frequency x-ray radiation \cite{Matsuoka2010,Kiselev2004,malka_rmp}. Apart from maximizing electron energy \cite{lu2006nonlinear,lu2007generating,lu2006nonlinear2}, the studies of LWFA can be important for increasing the total number of accelerated electrons, as well as intensity of the outgoing x-ray radiation. The latter can certainly benefit from highly nonlinear regimes at ultra-high intensities of the laser radiation. Motivated by this, in the present paper we will be analysing the effects 
of increasing laser intensities on the properties of the electrons and radiation spectra in LWFA.

In terms of potential applications, one of the main issues is the pulse depletion mechanism, which is of key importance for controlling the rate of energy transformation to other forms, as well as the proportion between them. The process of depletion is known to be difficult in terms of theoretical analysis and mathematical modelling even in the case of low intensities and densities \cite{Bulanov.pop.1992}. As a starting point, here we present a phenomenological analysis based on systematic comparison of simulations with and without taking into account the RR effect for different intensities of the driving laser pulse.

\onecolumngrid

\twocolumngrid

\subsection{Wakefield acceleration with radiation reaction included}
As we increase the intensity of the laser pulse generating the wake field, we expect that the electrons will be further accelerated. However, as the energy loss from synchrotron emission becomes appreciable, it will begin to affect the electron dynamics. This will then alter the characteristics of the wake field, particle acceleration process, and radiation emission.

When the RR force is properly modeled in a PIC simulation (in the classical regime), the dynamics of the LWFA changes. One of the main principal changes is that when the laser hits the plasma, not all electrons are forced around the laser pulse. Instead, some electrons can enter the high intensity part and start to co-propagate with the laser pulse, being trapped there by the so-called Radiation Reaction Trapping (RRT) effect \cite{RRT.prl.2014}. The number of such electrons is dependent on their energy, and thus the laser intensity. For the electrons trapped in the wakefield, the RR force reduces the acceleration of the electrons, and this also changes the spectrum of the emitted radiation. As the RR force acts like friction, we will also see pulse energy depletion at a faster rate than when not including this force. All-in-all, the combination of these, and other, effects makes for rather complex electron behaviour in an already nonlinear regime.

\section{Governing equations}
\subsection{The particle-in-cell method} 
Particle-in-cell (PIC) methods \cite{Birdsall1985,Dawson1983} are now a standard tool used for simulating laser-matter interactions. In this method the plasma is treated as an ensemble of particles moving in an electromagnetic (EM) field defined on a grid. The dynamics of the particles is calculated using the Lorentz force, and the resulting charge and current distributions in turn give rise to EM fields via Maxwell's equations. 

Even if the evolution of the plasma (and the laser) is solved self-consistently using the Lorentz force and Maxwell's equations in this model, there is still some important physics not included \cite{wallin2015, QEDPIC}. The grid and time resolutions place an upper limit on the frequency of the radiation accounted for in the simulations. This becomes a problem for relativistic particles, which can emit synchrotron radiation with a typical frequency of \cite{jackson1998classical}
\begin{equation}
  \omega_c = \frac{3}{2} \omega_H \gamma^3,
\end{equation}
where $\omega_H$ is the instantaneous cyclotron frequency of the particle and $\gamma$ the relativistic gamma factor. In a typical PIC simulation the particle motion (thus $\omega_H$) is resolved, but for a high enough $\gamma$ the emitted radiation, with typical frequency $\omega_c$, cannot be resolved.

This radiation is typically emitted over a small angle $\sim 1/\gamma$ and can be modelled as photons \cite{QEDPIC, wallin2015, PhysRevA.91.013822} instead of fields. As such these will have no direct impact on the particle dynamics of the simulation, but for relativistic particles the energy of this emitted radiation can constitute a large part of the particle energy. To accurately model this we must include the back reaction of the emitted radiation on the emitting particle, which we do via the RR force.

\subsection{The classical radiation reaction force} 
We include the effects of RR by adding a correction term to the equation of motion
\begin{equation}
  m \dot{\mathbf{v}} = \mathbf{F}_{\text{ext}} + \mathbf{F}_{\text{rad}},
\end{equation}
where the RR term is determined so that the work performed by the force is equal to the emitted energy. This produces the Abraham-Lorentz equation \cite{jackson1998classical} which contains a third order derivative, enabling unphysical runaway solutions. This can be avoided by approximating the third order time derivative with the Lorentz force, yielding the Landau-Lifshitz (LL) equation, valid when the RR force is much less than the Lorentz force in the instantaneous rest frame of the particle. The relativistic, covariant expression for the RR force is then given by \cite{LL.V2} 
\begin{eqnarray}
f^{\mu} &=& \frac{2e}{3}r_0 \partial_{\gamma} F^{\mu \nu} u_{\nu} u^{\gamma} + \notag \\
&+& \frac{2}{3} r_0^2 \Bigg[F^{\mu \alpha}F_{\alpha \nu} u^{\nu} + (F_{\nu \alpha} u^{\alpha}) (F^{\nu \beta} u_{\beta}) u^{\mu} \Bigg],
\end{eqnarray}
where $F^{\mu \nu}$ is the electromagnetic field tensor, $u_{\nu}$ the 4-velocity, and $r_0 = {e^2}/{m_e c^2}$ is the classical electron radius in cgs units. We neglect the first term since it's linear in the field strength, and its contribution is found to be negligible. (Indeed, it can be shown that, in cases where classical RR is important, the derivative term is even smaller than the electron spin force and so should be neglected out of consistency~\cite{Tamburini:2010}.) The three-dimensional form of the equation, without the first term, is \cite{LL.V2}
\begin{eqnarray}
  \mathbf{F}_{\text{rad}} &=& \frac{2}{3} r_0^2 \Bigg( \mathbf{E}\times \mathbf{B} + \frac{1}{c} \bigg[ \mathbf{B} \times (\mathbf{B}\times \mathbf{v} ) + (\mathbf{v} \cdot \mathbf{E}) \mathbf{E} \bigg] \notag \\
&-& \frac{\gamma^2}{c} \left[ \left( \mathbf{E} + \frac{1}{c} \mathbf{v} \times \mathbf{B}\right)^2 - \left( \frac{\mathbf{E} \cdot \mathbf{v}}{c}\right)^2 \right] \mathbf{v} \Bigg) .
\label{eq:RRForce_3D}
\end{eqnarray}
We note that although there are alternative formulations in the literature, all RR models result in approximately the same particle dynamics \cite{KravetsRR, VranicRR}, and the LL equation has shown to be consistent with QED \cite{Ilderton:2013tb}. For an overview of other models see Ref.~\cite{radiationreaction}.

\section{Simulations}
\begin{figure}
  \includegraphics[width=0.5\textwidth]{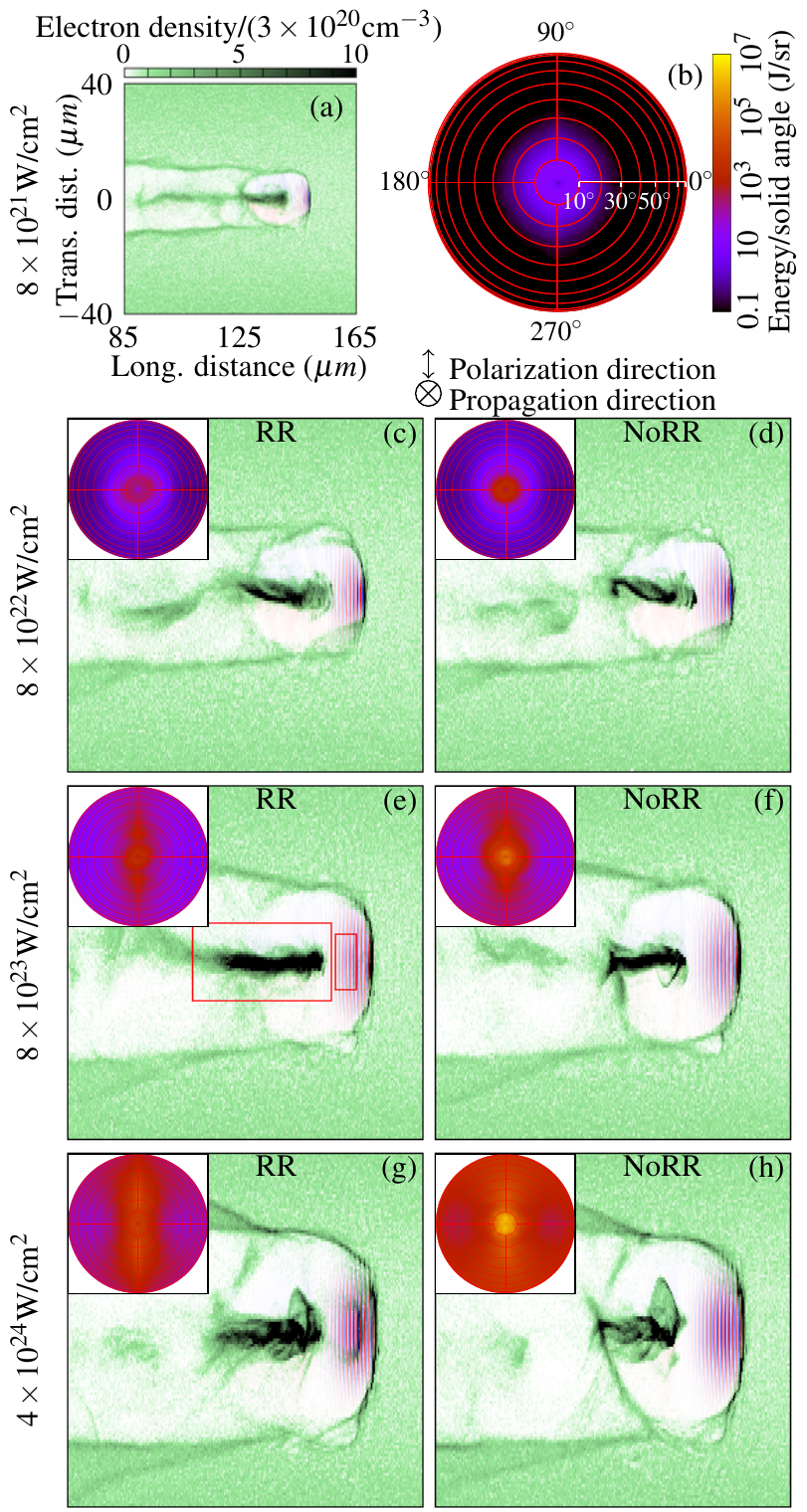}
  \caption{Comparison of the results for simulations with and without the RR force at $0.53$~ps, when the laser has propagated $160$ $\mu$m into the plasma. (a) The electron density (green to black) and the y-component of the electric field (red to blue) for the intensity equal to $8\times 10^{21}$W/cm$^2$, for which the two cases are inseparable. (b) The angular plot of the synchrotron emission for the same case. (c-h) The same information (Note: with the same scales and axes) for the cases with (left) and without (right) the RR force is shown in a more compact form for the intensities $8\times 10^{22}$W/cm$^2$, $8\times 10^{23}$W/cm$^2$ and $4\times 10^{24}$W/cm$^2$ counting from the top to the bottom. (e): Two boxes are showing the regions in which the amount of electrons in the laser pulse and electron bunch regions is measured.}
  \label{fig:DensityComparison+Angular}
\end{figure}

We use the 3D PIC code ELMIS \cite{gonoskov.phd.2013} with enabled modules for the classical RR in the form of the LL equation (\ref{eq:RRForce_3D}) and for the accounting of classical synchrotron emission via statistical routines \cite{wallin2015}. The modules are incorporated via the MDK (\emph{Module Development Kit}) interface \cite{QEDPIC}. This shared interface (with the PIC code PICADOR \cite{bastrakov.jcs.2012}) enables extensions of the PIC-scheme. The simulations are performed with a fixed plasma configuration for a number of different laser intensities, comparing the cases with and without the RR force. The laser pulse is a linearly polarized Gaussian shaped beam with duration $20 \, \mathrm{fs}$ and diameter $8 \, \mu\mathrm{m}$ (FWHM in intensity). The wavelength is $1 \, \mu\mathrm{m}$ and the pulse energy varies from $10^{2}\, \mathrm{J}$ to $5\times 10^{4}\, \mathrm{J}$ for different simulations, giving an intensity range from $8 \times 10^{21}\, \mathrm{W/cm}^2$ to $4 \times 10^{24}\, \mathrm{W/cm}^2$ (total energy over FWHM values of duration and diameter squared).

To enable highly nonlinear stable regimes for the whole range of intensity we consider for the target a near critical density hydrogen plasma with a radially (perpendicular to the pulse propagation direction) parabolic density profile. Electrons are initially inserted into the wakefield via a longitudinal (along the pulse propagation direction) density profile. The density of the plasma along the laser propagation axis is $N_{\text{axis}}=3 \times 10^{20}\, \mathrm{cm}^{-3}$, or $N_{\text{axis}} \approx 0.27 N_{\text{cr}}$ where $N_{\text{cr}}$ is the critical density, with the density increasing radially to $2 N_{\text{axis}}$ at $40 \, \mu\mathrm{m}$ (the edge of the simulation box). The longitudinal density profile of the plasma is such that the density increases from $0$ to $3N$ over the first $10 \, \mu\mathrm{m}$, then it decreases to $N$ over another $10 \, \mu\mathrm{m}$. This density is then kept for the remainder of the plasma. The size of the simulation box is $80 \, \mu\mathrm{m} \times 80\, \mu\mathrm{m} \times 80 \, \mu\mathrm{m}$, on $512 \times 128 \times 128$ cells with $2$ macroparticles per cell. The simulation box is comoving with the laser pulse, though the simulation is still performed in the lab frame. In Fig. \ref{fig:DensityComparison+Angular} pictures of the simulations after $0.53 \, \mathrm{ps}$ can be seen, with the intensity ranging from  $8 \times 10^{21}\, \mathrm{W/cm}^2$ to $4 \times 10^{24}\, \mathrm{W/cm}^2$. 

The high density (near critical) plasma results in a high level of interaction between the particles and the pulse, depleting the pulse after a moderate amount of time. For the laser intensity of $4 \times 10^{23} \, \mathrm{W/cm}^2$ the pulse loses half of its energy in $1.7 \, \mathrm{ps}$, having travelled $L \approx 500 \lambda$. This enables us to study the effect of the pulse depletion, for not too long simulation times.

\section{Results}
\begin{figure}
\begin{tikzpicture}
  \node[anchor=south west,inner sep=0] at (0,0)
       {\includegraphics[width=66mm]{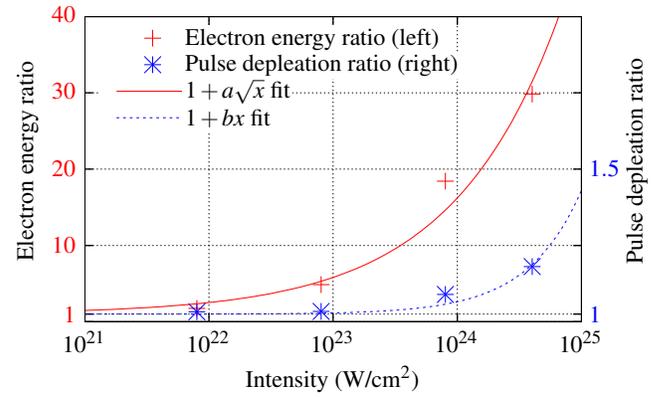}};

\pgfmathsetmacro\w{6.6} \pgfmathsetmacro\h{4.1}

  \node [below] at (0,0) {$10^{21}$}; \node [below] at (1.65,0)
        {$10^{22}$}; \node [below] at (1.65*2,0) {$10^{23}$}; \node
        [below] at (1.65*3,0) {$10^{24}$}; \node [below] at (1.65*4,0)
        {$10^{25}$};

  \foreach \x in {1,10,20,30,40} \node [left,color=red] at
  (0,\x*\h/40) {$\x$};

  \node [right,color=blue] at (6.6,\h/40*1) {$1$}; \node
        [right,color=blue] at (6.6,\h/40*20) {$1.5$};

  \node [rotate=90] at (-0.75,2.15) {Electron energy ratio}; 
  \node [rotate=90] at (6.6+0.75,2.15) {Pulse depleation ratio}; 
  \node at (3.3,-0.75) {Intensity (W/cm$^2$)};

  \node [right,fill=white, inner sep=1pt] at (1.3,3.77) {Electron energy ratio (left)}; 
  \node [right,fill=white, inner sep=1pt] at (1.3,3.77-0.35*1) {Pulse depleation ratio (right)};
  \node [right,fill=white, inner sep=1pt] at (1.3,3.77-0.35*3) {$1+bx$ fit}; 
  \node [right,fill=white, inner sep=1pt] at (1.3,3.77-0.35*2) {$1+a\sqrt{x}$ fit};

\end{tikzpicture}

  \caption{Ratio of peak electron energies (left axis) and pulse depletion times (right axis) for simulations without to those with RR included. The pulse depletion time is calculated as the time taken for the laser to lose half its initial energy. The ratio of the peak electron energies in the two cases is calculated throughout the simulations and is generally found to increase with time. Here we plot the maximum value which typically occurs when the pulse is depleted. Note the different scales on the vertical axes.}
  \label{fig:Combined}
\end{figure}

\begin{figure}
\begin{tikzpicture}
  \node[anchor=south west,inner sep=0] (pic) at (0,0) {\includegraphics[width=72mm]{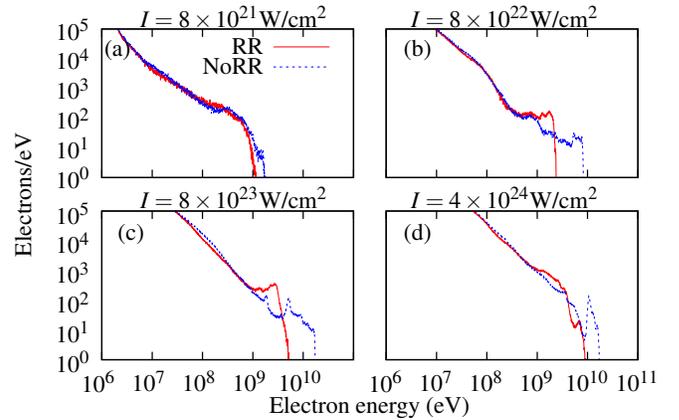}};

\pgfmathsetmacro\w{7.2}
\pgfmathsetmacro\h{7.2*0.618034}

\node [below] at (-\w*0.465*6/5+\w*0.465*10/5,0) {$10^{10}$};
\node [below] at (-\w*0.465*6/5+\w*0.465*9/5,0) {$10^{9}$};
\node [below] at (-\w*0.465*6/5+\w*0.465*8/5,0) {$10^{8}$};
\node [below] at (-\w*0.465*6/5+\w*0.465*7/5,0) {$10^{7}$};
\node [below] at (-\w*0.465*6/5+\w*0.465*6/5,0) {$10^{6}$};

\node [below] at (0.53*\w-\w*0.465*6/5+\w*0.465*11/5,0) {$10^{11}$};
\node [below] at (0.53*\w-\w*0.465*6/5+\w*0.465*10/5,0) {$10^{10}$};
\node [below] at (0.53*\w-\w*0.465*6/5+\w*0.465*9/5,0) {$10^{9}$};
\node [below] at (0.53*\w-\w*0.465*6/5+\w*0.465*8/5,0) {$10^{8}$};
\node [below] at (0.53*\w-\w*0.465*6/5+\w*0.465*7/5,0) {$10^{7}$};
\node [below] at (0.53*\w-\w*0.465*6/5+\w*0.465*6/5,0) {$10^{6}$};

\node [left] at (0,\h*0.55+\h*0.445/5*0) {$10^{0}$};
\node [left] at (0,\h*0.55+\h*0.445/5*1) {$10^{1}$};
\node [left] at (0,\h*0.55+\h*0.445/5*2) {$10^{2}$};
\node [left] at (0,\h*0.55+\h*0.445/5*3) {$10^{3}$};
\node [left] at (0,\h*0.55+\h*0.445/5*4) {$10^{4}$};
\node [left] at (0,\h*0.55+\h*0.445/5*5) {$10^{5}$};

\node [left] at (0,\h*0.445/5*0) {$10^{0}$};
\node [left] at (0,\h*0.445/5*1) {$10^{1}$};
\node [left] at (0,\h*0.445/5*2) {$10^{2}$};
\node [left] at (0,\h*0.445/5*3) {$10^{3}$};
\node [left] at (0,\h*0.445/5*4) {$10^{4}$};
\node [left] at (0,\h*0.445/5*5) {$10^{5}$};

\node [rotate=90] at (-1.0,\h/2) {Electrons/eV};
\node at (\w/2,-0.6) {Electron energy (eV)};

\node [left] at (\w*0.32,\h*0.945) {RR};
\node [left] at (\w*0.32,\h*0.945-0.06*\h) {NoRR};

\node [above] at (\w/4,0.97*\h) {$I=8\times 10^{21}$W/cm$^2$};
\node [above] at (3*\w/4,0.97*\h) {$I=8\times 10^{22}$W/cm$^2$};
\node [above] at (\w/4,0.43*\h) {$I=8\times 10^{23}$W/cm$^2$};
\node [above] at (3*\w/4,0.43*\h) {$I=4\times 10^{24}$W/cm$^2$};

\node [below left] at (0.075*\w,0.99*\h) {(a)};
\node [below left] at (0.63*\w,0.99*\h) {(b)};
\node [below left] at (0.1*\w,\h*0.44) {(c)};
\node [below left] at (0.63*\w,\h*0.44) {(d)};
\end{tikzpicture}
  \caption{Electron energy distribution at $1.06$ ps. The inclusion of RR reduces the peak energy of the electrons. }
  \label{fig:electron_energy_distribution}
\end{figure}

\begin{figure}
\includegraphics[width=0.5\textwidth]{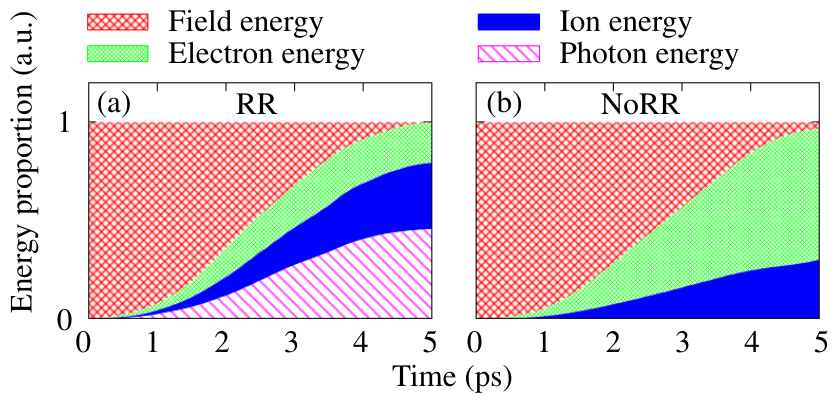}
  \caption{Redistribution of laser pulse energy to electrons, ions (hydrogen) and photons over time for the intensity $4\times 10^{24}$W/cm$^2$. The energy of the electrons is clearly overestimated without RR. Also, a substantial amount of energy goes to the ions in both cases. }
  \label{fig:EnergyChannels}
\end{figure}

We compare, for a number of different intensities, how the inclusion of the RR force affects the time it takes for the laser pulse to be depleted. The result can be seen in Fig.~\ref{fig:Combined}, where we plot the ratio of the pulse depletion times (i.e.~the time for the pulse to lose half its energy) without and with the RR force included ($T_{\text{NoRR}}/T_{\text{RR}}$).  It can be seen that when the laser intensity surpasses $10^{23}$ W/cm$^2$ this ratio begins to differ from $1$, indicating that RR effects are starting to reduce the propagation distance of the laser. If the pulse loses too much of its energy this effect could prevent the application of wakefield acceleration schemes where long plasmas are used. The inclusion of the RR force is equivalent to restoring the possibility of the electrons emitting high energy photons. Thus the faster pulse depletion can be understood as being a result of the extra work the field does in re-accelerating these electrons. This decrease in energy of the electrons can also be seen in Fig.~\ref{fig:Combined}, where the ratio of the maximum electron energies without and with RR included is shown. For intensities above $10^{22}$ W/cm$^2$ this starts to be noticeable and for even higher intensities there is a great difference in the maximum electron energy due to the inclusion of RR. An example of the electron energy distribution, after $1.06$ ps can be seen in Fig. \ref{fig:electron_energy_distribution}. In Fig. \ref{fig:EnergyChannels} an example of the transformation of the initial pulse energy to different types of particle energy can be seen, for the two cases with and without RR for the intensity $4 \times 10^{24}$ W/cm$^2$. It is shown that the pulse is drained faster in the case where RR is included, and one can also see how the total electron energy is overestimated when it is not. Also, one can see that a great amount of energy actually goes to the ions. This is regardless of whether RR is included or not.

Our regime of simulation is similar to LWFA, but with some important differences. The field intensity is sufficiently strong to perturb the hydrogen ions, and their motion changes the dynamics of the laser-plasma interaction \cite{capdessus2013influence}. One effect of this is that the particle densities do not close behind the bubble as in conventional LWFA; instead there is a column after the bubble region containing substantially fewer particles, both electrons and ions. There seem to be two effects on the ions as they are hit by the laser, (i) some ions are forced around the pulse by the radiation pressure and (ii) some ions pass through the laser pulse region and interact strongly with the electron bunch. The result is a column behind the bubble region void of ions, except a filament in the center. The void region does not attract either particle species and the particle densities do not close behind the bubble. In spite of this, the bubble is still formed and electrons stream in and become trapped behind the laser just as in conventional LWFA, as seen in Fig. \ref{fig:DensityComparison+Angular}. The ion density in the laser pulse region is relatively unchanged and there is thus still a strong electric field, induced by the charge separation, which can accelerate the electrons. 

\begin{figure}
\begin{tikzpicture}
  \node[anchor=south west,inner sep=0] (pic) at (0,0) {\includegraphics[width=75mm]{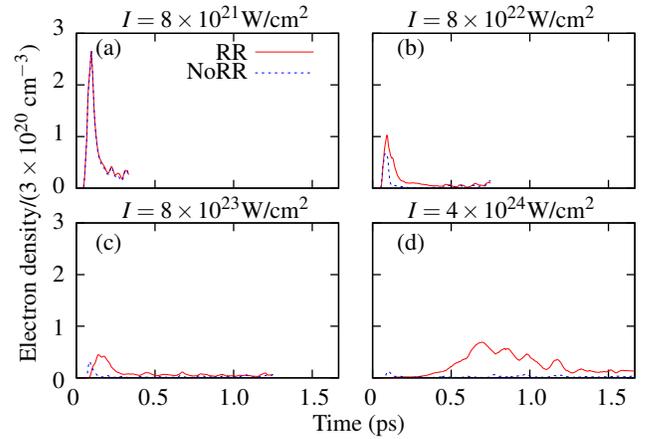}};

\pgfmathsetmacro\w{7.6}
\pgfmathsetmacro\h{4.7}

\foreach \x in {0,0.5,1.0,1.5} \node [below] at (\x*\w*0.47/1.67,0) {$\x$};
\foreach \x in {0,0.5,1.0,1.5} \node [below] at (\w*0.52+\x*\w*0.47/1.67,0) {$\x$};

\foreach \x in {0,1,2,3} \node [left] at (0,\h*0.45/3*\x) {$\x$};
\foreach \x in {0,1,2,3} \node [left] at (0,\h*0.55+\h*0.45/3*\x) {$\x$};

\node [rotate=90] at (-0.6,\h/2) {Electron density/($3\times 10^{20}$ cm$^{-3}$)};
\node at (\w/2,-0.6) {Time (ps)};

\node [left] at (\w*0.32,\h*0.93) {RR};
\node [left] at (\w*0.32,\h*0.93-0.26) {NoRR};

\node [above] at (\w/4,0.97*\h) {$I=8\times 10^{21}$W/cm$^2$};
\node [above] at (3*\w/4,0.97*\h) {$I=8\times 10^{22}$W/cm$^2$};
\node [above] at (\w/4,0.43*\h) {$I=8\times 10^{23}$W/cm$^2$};
\node [above] at (3*\w/4,0.43*\h) {$I=4\times 10^{24}$W/cm$^2$};

\node [below left] at (0.1*\w,0.99*\h) {(a)};
\node [below left] at (0.63*\w,0.99*\h) {(b)};
\node [below left] at (0.1*\w,\h*0.44) {(c)};
\node [below left] at (0.63*\w,\h*0.44) {(d)};

\end{tikzpicture}
  \caption{Electron density in the laser pulse region, measured in a box as shown in Fig. \ref{fig:DensityComparison+Angular}(e). The range shown is until the laser pulse starts to break down and the wakefield collapses allowing particles flow freely through the box. For the lowest intensity, the electrons don't flow as much around the pulse, but pass through it and there is no difference between the RR and no RR cases. For higher intensities the electrons are forced around the pulse, but the effect of RR counteracts this. More electrons flow through the pulse due to RR, and there is also a trapping effect, where for the higher intensity cases the electrons spend more time in the high intensity region. The thickness of the box is $2 \mu m$, the same as the electron density slice in Fig. \ref{fig:DensityComparison+Angular}}
  \label{fig:ChargeOverTimeMulti}
\end{figure}
\begin{figure}
  \begin{tikzpicture}
    \node[anchor=south west,inner sep=0] (pic) at (0,0) {\includegraphics[width=75mm]{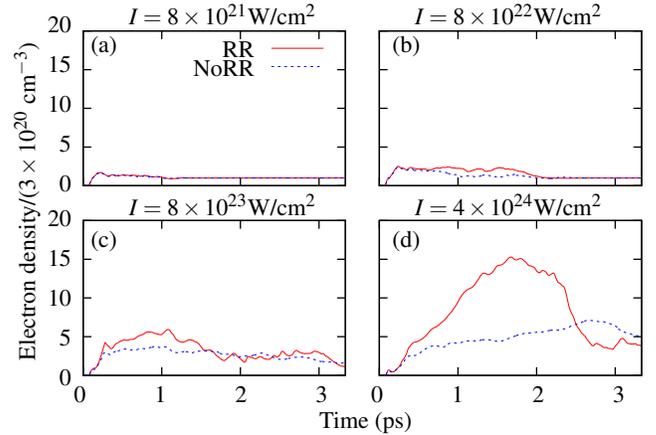}};

    \pgfmathsetmacro\w{7.6}
    \pgfmathsetmacro\h{4.7}

    \foreach \x in {0,1,2,3} \node [below] at (\x*\w*0.47/3.335,0) {$\x$};
    \foreach \x in {0,1,2,3} \node [below] at (\w*0.52+\x*\w*0.47/3.335,0) {$\x$};

    \foreach \x in {0,5,10,15,20} \node [left] at (0,\h*0.45/20*\x) {$\x$};
    \foreach \x in {0,5,10,15,20} \node [left] at (0,\h*0.55+\h*0.45/20*\x) {$\x$};

    \node [rotate=90] at (-0.7,\h/2) {Electron density/($3\times 10^{20}$ cm$^{-3}$)};
    \node at (\w/2,-0.6) {Time (ps)};

    \node [left] at (\w*0.32,\h*0.93) {RR};
    \node [left] at (\w*0.32,\h*0.93-0.26) {NoRR};

    \node [above] at (\w/4,0.97*\h) {$I=8\times 10^{21}$W/cm$^2$};
    \node [above] at (3*\w/4,0.97*\h) {$I=8\times 10^{22}$W/cm$^2$};
    \node [above] at (\w/4,0.43*\h) {$I=8\times 10^{23}$W/cm$^2$};
    \node [above] at (3*\w/4,0.43*\h) {$I=4\times 10^{24}$W/cm$^2$};

    \node [below left] at (0.08*\w,0.99*\h) {(a)};
    \node [below left] at (0.61*\w,0.99*\h) {(b)};
    \node [below left] at (0.08*\w,\h*0.44) {(c)};
    \node [below left] at (0.61*\w,\h*0.44) {(d)};
  \end{tikzpicture}

  \caption{Electron density in the electron bunch region, measured in a box around the bunch as shown in Fig. \ref{fig:DensityComparison+Angular}(e). 
  At higher intensities the inclusion of RR increases the amount of charge in the bunch up to the point were the laser is too depleted to drive the bunch.
  For the case without RR the bunch electrons, as well as the laser pulse, have lost less energy and the bunch can be sustained for a longer time. For the lower intensities the pulse is quickly depleted, after which only the background plasma is measured.}
  \label{fig:chargeOverTimeInBunch}
\end{figure}

The inclusion of the RR force also changes the interaction between the laser pulse and the plasma. One such effect is that electrons can become trapped and pass through the high intensity pulse region of the laser as it enters the plasma. This can be seen in Fig. \ref{fig:ChargeOverTimeMulti} where the electron density in the high intensity part of the pulse is plotted as a function of time. For the lower intensity there is no visible effect of the RR. Some electrons pass through the pulse region as this is not sufficiently strong to force them around. As the laser gets more intense the electrons are forced around the pulse, and the electrons in the high intensity region decrease. The effect of including RR is to change this, allowing electrons to go through the high intensity pulse region, as well as to spending a longer time there. In Fig. \ref{fig:DensityComparison+Angular} the electron density and the laser pulse is shown for the different intensity cases. The trapped particles passing through the intense pulse region can be seen for the intense cases when RR is included (Fig. \ref{fig:DensityComparison+Angular}(g)). One can also note the difference in the size of the electron bunch, with a larger bunch for the cases when RR is included. This is further shown in Fig. \ref{fig:chargeOverTimeInBunch} where electron density in the bubble region is plotted as a function of time.

\begin{figure}
\begin{tikzpicture}
  \node[anchor=south west,inner sep=0] (pic) at (0,0) {\includegraphics[width=72mm]{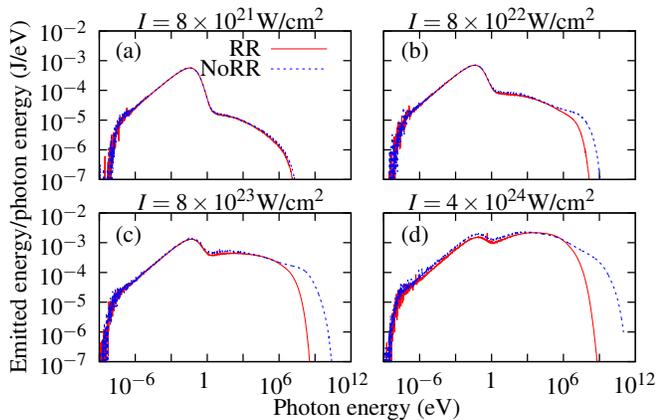}};

\pgfmathsetmacro\w{7.2}
\pgfmathsetmacro\h{7.2*0.618034}

\node [below] at (0.53*\w+\w*0.465*9/21+\w*0.465*12/21,0) {$10^{12}$};
\node [below] at (0.53*\w+\w*0.465*9/21+\w*0.465*6/21,0) {$10^{6}$};
\node [below] at (0.53*\w+\w*0.465*9/21+\w*0.465*0/21,0) {$1$};
\node [below] at (0.53*\w+\w*0.465*9/21-\w*0.465*6/21,0) {$10^{{-}6}$};

\node [below] at (\w*0.465*9/21+\w*0.465*12/21,0) {$10^{12}$};
\node [below] at (\w*0.465*9/21+\w*0.465*6/21,0) {$10^{6}$};
\node [below] at (\w*0.465*9/21+\w*0.465*0/21,0) {$1$};
\node [below] at (\w*0.465*9/21-\w*0.465*6/21,0) {$10^{{-}6}$};

\node [left] at (0,0) {$10^{-7}$};
\node [left] at (0,\h*0.445/5*1) {$10^{-6}$};
\node [left] at (0,\h*0.445/5*2) {$10^{-5}$};
\node [left] at (0,\h*0.445/5*3) {$10^{-4}$};
\node [left] at (0,\h*0.445/5*4) {$10^{-3}$};
\node [left] at (0,\h*0.445/5*5) {$10^{-2}$};

\node [left] at (0,\h*0.55+\h*0.445/5*0) {$10^{-7}$};
\node [left] at (0,\h*0.55+\h*0.445/5*1) {$10^{-6}$};
\node [left] at (0,\h*0.55+\h*0.445/5*2) {$10^{-5}$};
\node [left] at (0,\h*0.55+\h*0.445/5*3) {$10^{-4}$};
\node [left] at (0,\h*0.55+\h*0.445/5*4) {$10^{-3}$};
\node [left] at (0,\h*0.55+\h*0.445/5*5) {$10^{-2}$};

\node [rotate=90] at (-1.0,\h/2) {Emitted energy/photon energy (J/eV)};
\node at (\w/2,-0.6) {Photon energy (eV)};

\node [left] at (\w*0.32,\h*0.945) {RR};
\node [left] at (\w*0.32,\h*0.945-0.06*\h) {NoRR};

\node [above] at (\w/4,0.97*\h) {$I=8\times 10^{21}$W/cm$^2$};
\node [above] at (3*\w/4,0.97*\h) {$I=8\times 10^{22}$W/cm$^2$};
\node [above] at (\w/4,0.43*\h) {$I=8\times 10^{23}$W/cm$^2$};
\node [above] at (3*\w/4,0.43*\h) {$I=4\times 10^{24}$W/cm$^2$};

\node [below left] at (0.1*\w,0.99*\h) {(a)};
\node [below left] at (0.63*\w,0.99*\h) {(b)};
\node [below left] at (0.1*\w,\h*0.44) {(c)};
\node [below left] at (0.63*\w,\h*0.44) {(d)};

\end{tikzpicture}
  \caption{Radiation spectra for the four different intensity cases. For the lowest intensity there is little difference between the RR and no RR cases, but this grows very notable for the higher intensities where the inclusion of RR removes the high frequency contribution, as the peak energy of the electrons is lowered.}
  \label{fig:radiationSpectra}
\end{figure}

\begin{figure}
\begin{tikzpicture}
  \node[anchor=south west,inner sep=0] (pic) at (0,0) {\includegraphics[width=75mm]{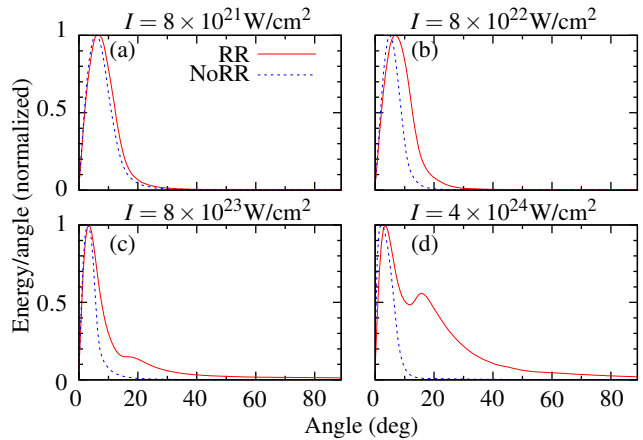}};

\pgfmathsetmacro\w{7.6}
\pgfmathsetmacro\h{4.7}

\foreach \x in {0,20,40,60,80} \node [below] at (\x*\w*0.47/90,0) {$\x$};
\foreach \x in {0,20,40,60,80} \node [below] at (0.52*\w+\x*\w*0.47/90,0) {$\x$};

\foreach \x in {0,0.5,1} \node [left] at (0,\h*0.45/1*\x) {$\x$};
\foreach \x in {0,0.5,1} \node [left] at (0,\h*0.55+\h*0.45/1*\x) {$\x$};

\node [rotate=90] at (-0.7,\h/2) {Energy/angle (normalized)};
\node at (\w/2,-0.6) {Angle (deg)};

\node [left] at (\w*0.32,\h*0.93) {RR};
\node [left] at (\w*0.32,\h*0.93-0.26) {NoRR};

\node [above] at (\w/4,0.97*\h) {$I=8\times 10^{21}$W/cm$^2$};
\node [above] at (3*\w/4,0.97*\h) {$I=8\times 10^{22}$W/cm$^2$};
\node [above] at (\w/4,0.43*\h) {$I=8\times 10^{23}$W/cm$^2$};
\node [above] at (3*\w/4,0.43*\h) {$I=4\times 10^{24}$W/cm$^2$};

\node [below left] at (0.04*\w+0.08*\w,0.99*\h) {(a)};
\node [below left] at (0.04*\w+0.61*\w,0.99*\h) {(b)};
\node [below left] at (0.04*\w+0.08*\w,\h*0.44) {(c)};
\node [below left] at (0.04*\w+0.61*\w,\h*0.44) {(d)};

\end{tikzpicture}

  \caption{Angular distribution of radiated energy as a function of the angle from the propagation direction of the laser pulse. Note that the energy scale is normalized.}
  \label{fig:AngularPlot1FNormalized}
\end{figure}

The photon emission calculated from the electron motion is vastly affected by the inclusion of the RR force for these intensities, as expected. In Fig. \ref{fig:radiationSpectra} the frequency spectra of the emitted photon radiation is shown. The inclusion of RR prevents the very highest electron energies, and their high frequency contribution is removed from the spectra. This is also visible in Figs. \ref{fig:DensityComparison+Angular} and \ref{fig:AngularPlot1FNormalized} where the angular spectra of the emitted radiation is shown. The high frequency radiation removed by the inclusion of RR was emitted very close to the propagation direction of the pulse, which can be seen in the broader spectra for the case with RR included.

\section{Conclusions}
Including a radiation reaction force in the equation of motion for electrons in a PIC simulation of a laser induced plasma wakefield can alter the dynamics of the electrons. Our simulations show that it can affect the pulse depletion distance for laser intensities from $10^{23}\, \mathrm{W/cm}^2$ and the energy of the electrons in the electron bunch from intensities of $10^{22}\, \mathrm{W/cm}^2$, as well as affecting how the electrons interact with the laser. 

The RR force models the radiation from the electrons which is of too high frequency/short wavelength (thus of large energy) to be resolved by the timestep and the grid. The inclusion of this restores an important channel of energy loss for the electrons, with the result that the electrons in the bunch become less energetic than they would otherwise be. There is then also an effect where the electrons take more energy from the laser pulse, shortening the pulse depletion distance. The energy lost due to the RR force is emitted as photons. In the considered cases these have little impact on the particle dynamics, but allow for a wakefield setup to be used as a gamma radiation source.

Our simulations reveal a highly nonlinear regime of laser pulse propagation in underdense or near critical plasmas, similar to LWFA but with  sufficiently high intensity for ion motion to be of importance. This affect the dynamics of the laser-plasma interaction, but still allows for electron acceleration.

For intensities approaching $10^{24} \, \mathrm{W/cm}^2$ and beyond, it is necessary to approach the photon emission and RR using QED methods \cite{dipiazza-rmp}. We have implemented such a stochastic QED-based routine \cite{QEDPIC} for emission of photons and also pair production, and used for comparison for intensities up to $4 \times 10^{24}\, \mathrm{W/cm}^2$. The results are similar to the case of the classical RR force as presented here and the main results are unchanged, even if the QED runs indicate that the energy loss of the electrons is slightly overestimated by the classical RR force.

For even higher intensities than those presented here, the stochastic nature of quantum photon emission, and the resulting electron recoil is likely to change the dynamics of the wakefield significantly. In this case, due to the large number of hard photons emitted, Breit--Wheeler pair production will also have to be included \cite{QEDPIC}. However, we leave this issue for future studies.

\section{Method} \label{sec:Method}
The RR force is included into the PIC codes \emph{ELMIS3D} and \emph{Picador} \cite{QEDPIC, bastrakov.jcs.2012} via Eq.~(\ref{eq:RRForce_3D}). For each electron (super)particle and timestep, the electric and magnetic fields are weighted to the position of the particle from the nearby grid cells. These are used, together with the particle gamma-factor and velocity to calculate the RR force on the particle. The force then affects the momentum of the particle as 
\begin{equation}
  \Delta \mathbf{p} = \mathbf{F} \cdot \Delta t,
\end{equation}
where $\Delta t$ is the timestep, and this is then added to the particle momentum. Then the particle is affected by the Lorentz force via the \emph{Boris scheme} \cite{boris1970relativistic}, as is customary in PIC schemes.

The added RR force represent the energy lost to radiation, both high energy radiation not resolvable by the grid and coherent emission accounted for by the grid. There is thus a double counting for the low frequency, coherent radiation. However, the RR force for the case of coherent emission has only a negligible effect on the particle dynamics as the energy lost via this radiation is much less than the particle energy \cite{QEDPIC}.

The RR force is only calculated for the electrons. This is a good approximation, as the much lighter electrons are accelerated to higher energies, and thus are more affected by the RR force. 

Using a RR force is a good approximation up to laser intensities where QED begin to be of importance, and the  electrons can radiate a large amount of their energy in a single photon emission \cite{QEDPIC}. Then the photon emission is more stochastic and the averaged energy loss due to the RR force is not applicable.

\begin{acknowledgements}
This research was supported by the Swedish Research Council Grants \# 2013-4248 and 2012-5644, and the Knut \& Alice Wallenberg Foundation Grant \textit{Plasma based compact ion sources}. The simulations were performed on resources provided by the Swedish National Infrastructure for Computing (SNIC)  at High Performance Computing Center North (HPC2N). A.G. also acknowledges the Russian Foundation for Basic Research, project No. 15-37-21015.
\end{acknowledgements}

\bibliography{References}

\begin{thebibliography}{10}

\bibitem{tajima-dawson}
T.~Tajima and J.M. Dawson.
\newblock Laser electron accelerator.
\newblock {\em Phys. Rev. Lett.}, 43(4):267--270, 1979.

\bibitem{malka_rmp}
S.~Corde, K.~Ta~Phuoc, G.~Lambert, R.~Fitour, V.~Malka, A.~Rousse, A.~Beck, and
  E.~Lefebvre.
\newblock Femtosecond x-rays from laser-plasma accelerators.
\newblock {\em Rev. Mod. Phys.}, 85(1):1--48, 2013.

\bibitem{malka2008principles}
Victor Malka, J{\'e}r{\^o}me Faure, Yann~A Gauduel, Erik Lefebvre, Antoine
  Rousse, and Kim~Ta Phuoc.
\newblock Principles and applications of compact laser--plasma accelerators.
\newblock {\em Nature Phys.}, 4(6):447--453, 2008.

\bibitem{vulcan}
Vulcan: www.clf.stfc.ac.uk.

\bibitem{eli}
ELI: www.extreme-light infrastructure.eu.

\bibitem{xcels}
XCELS: www.xcels.iapras.ru.

\bibitem{mourou-rmp}
G.A. Mourou, T.~Tajima, and S.V. Bulanov.
\newblock Optics in the relativistic regime.
\newblock {\em Rev. Mod. Phys.}, 78(2):309--371, 2006.

\bibitem{marklund-shukla}
M.~Marklund and P.K. Shukla.
\newblock Nonlinear collective effects in photon-photon and photon-plasma
  interactions.
\newblock {\em Rev. Mod. Phys.}, 78(2):591--640, 2006.

\bibitem{Heinzl:2008an}
Thomas Heinzl and Anton Ilderton.
\newblock {Exploring high-intensity QED at ELI}.
\newblock {\em Eur. Phys. J.}, D55:359--364, 2009.

\bibitem{dipiazza-rmp}
A.~Di~Piazza, C.~M\"uller, K.Z. Hatsagortsyan, and C.H. Keitel.
\newblock Extremely high-intensity laser interactions with fundamental quantum
  systems.
\newblock {\em Rev. Mod. Phys.}, 84(3):1177--1228, 2012.

\bibitem{VranicRR}
M.~{Vranic}, J.~L. {Martins}, R.~A. {Fonseca}, and L.~O. {Silva}.
\newblock {Classical Radiation Reaction in Particle-In-Cell Simulations}.
\newblock {\em ArXiv e-prints 1502.02432}, February 2015.

\bibitem{radiationreaction}
D.A. Burton and Noble. A.
\newblock Aspects of electromagnetic radiation reaction in strong fields.
\newblock {\em Contemporary Physics}, 55(2):110--121, 2014.

\bibitem{wallin2015}
E.~Wallin, A.~Gonoskov, and M.~Marklund.
\newblock Effects of high energy photon emissions in laser generated
  ultra-relativistic plasmas: real-time synchrotron simulations.
\newblock {\em Phys. Plasmas}, 22:033117, 2015.

\bibitem{QEDPIC}
A.~Gonoskov, S.~Bastrakov, E.~Efimenko, A.~Ilderton, M.~Marklund, I.~Meyerov,
  A.~Muraviev, A.~Sergeev, I.~Surmin, and E.~Wallin.
\newblock Extended particle-in-cell schemes for physics in ultrastrong laser
  fields: Review and developments.
\newblock {\em Phys. Rev. E}, 92:023305, Aug 2015.

\bibitem{Sprangle1988}
P.~Sprangle, E.~Esarey, a.~Ting, and G.~Joyce.
\newblock {Laser wakefield acceleration and relativistic optical guiding}.
\newblock {\em Appl. Phys. Lett.}, 53(22):2146, 1988.

\bibitem{Leemans2002}
W.~Leemans, P.~Catravas, E.~Esarey, C.~Geddes, C.~Toth, R.~Trines,
  C.~Schroeder, B.~Shadwick, J.~van Tilborg, and J.~Faure.
\newblock {Electron-Yield Enhancement in a Laser-Wakefield Accelerator Driven
  by Asymmetric Laser Pulses}.
\newblock {\em Phys. Rev. Lett.}, 89(17):174802, October 2002.

\bibitem{Malka2002}
V.~Malka, S.~Fritzler, E.~Lefebvre, and M.M. Aleonard.
\newblock {Electron acceleration by a wake field forced by an intense
  ultrashort laser pulse}.
\newblock {\em Science}, 298(November):1596--1601, 2002.

\bibitem{Modena1995}
A.~Modena, Z.~Najmudin, A.~Dangor, E., C.~E. Clayton, K.~A. Marsh, C.~Joshi,
  V.~Malka, C.~B. Darrow, C.~Danson, D.~Neely, and F.~N. Walsh.
\newblock {Electron acceleration from the breaking of relativistic plasma
  waves}.
\newblock {\em Nature}, 377, 1995.

\bibitem{Kiselev2004}
S.~Kiselev, A.~Pukhov, and I.~Kostyukov.
\newblock {X-ray Generation in Strongly Nonlinear Plasma Waves}.
\newblock {\em Phys. Rev. Lett.}, 93(13):135004, September 2004.

\bibitem{Geddes2004}
C.G.R. Geddes, C.~Toth, and J.~Van Tilborg.
\newblock {High-quality electron beams from a laser wakefield accelerator using
  plasma-channel guiding}.
\newblock {\em Nature}, 431(September), 2004.

\bibitem{leemans2006gev}
W.P. Leemans, B.~Nagler, A.J. Gonsalves, C.~Toth, K.~Nakamura, C.G.R. Geddes,
  E.~Esarey, C.B. Schroeder, and S.M. Hooker.
\newblock {GeV electron beams from a centimetre-scale accelerator}.
\newblock {\em Nature Phys.}, 2(10):696--699, 2006.

\bibitem{Matsuoka2010}
T.~Matsuoka, S.~Kneip, C.~McGuffey, C.~Palmer, J.~Schreiber, C.~Huntington,
  Y.~Horovitz, F.~Dollar, V.~Chvykov, G.~Kalintchenko, A.G.R. Thomas,
  V.~Yanovsky, K.~Ta Phuoc, S.P.D. Mangles, Z.~Najmudin, A.~Maksimchuk, and
  K.~Krushelnick.
\newblock {Synchrotron x-ray radiation from laser wakefield accelerated
  electron beams in a plasma channel}.
\newblock {\em J.Phys. Conf. Ser.}, 244(4):042026, August 2010.

\bibitem{lu2006nonlinear}
Wei Lu, Chengkun Huang, Miaomiao Zhou, WB~Mori, and T~Katsouleas.
\newblock Nonlinear theory for relativistic plasma wakefields in the blowout
  regime.
\newblock {\em Phys. Rev. Lett.}, 96(16):165002, 2006.

\bibitem{lu2007generating}
Wei Lu, M~Tzoufras, C~Joshi, FS~Tsung, WB~Mori, J~Vieira, RA~Fonseca, and
  LO~Silva.
\newblock {Generating multi-GeV electron bunches using single stage laser
  wakefield acceleration in a 3D nonlinear regime}.
\newblock {\em Phys. Rev. ST Accel. Beams}, 10(6):061301, 2007.

\bibitem{lu2006nonlinear2}
W~Lu, C~Huang, M~Zhou, M~Tzoufras, FS~Tsung, WB~Mori, and T~Katsouleas.
\newblock A nonlinear theory for multidimensional relativistic plasma wave
  wakefieldsa).
\newblock {\em Phys. Plasmas}, 13(5):056709, 2006.

\bibitem{Bulanov.pop.1992}
S.V. Bulanov, I.N. Inovenkov, V.I. Kirsanov, N.M. Naumova, and A.S. Sakharov.
\newblock Nonlinear depletion of ultrashort and relativistically strong laser
  pulses in an underdense plasma.
\newblock {\em Phys. Fluids B: Plasma Phys.}, 4(7):1935--1942, 1992.

\bibitem{RRT.prl.2014}
L.~L. Ji, A.~Pukhov, I.~Yu. Kostyukov, B.~F. Shen, and K.~Akli.
\newblock Radiation-reaction trapping of electrons in extreme laser fields.
\newblock {\em Phys. Rev. Lett.}, 112:145003, Apr 2014.

\bibitem{Birdsall1985}
C.K. Birdsall and A.B. Langdon.
\newblock Plasma physics via computer simulation.
\newblock 1985.

\bibitem{Dawson1983}
J.M. Dawson.
\newblock {Particle simulation of plasmas}.
\newblock {\em Rev.Mod. Phys.}, 1983.

\bibitem{jackson1998classical}
J.D. Jackson.
\newblock Classical electrodynamics.
\newblock {\em Classical Electrodynamics, 3rd Edition, by John David Jackson,
  pp. 832. ISBN 0-471-30932-X. Wiley-VCH, July 1998.}, 1, 1998.

\bibitem{PhysRevA.91.013822}
C.~N. Harvey, A.~Ilderton, and B.~King.
\newblock Testing numerical implementations of strong-field electrodynamics.
\newblock {\em Phys. Rev. A}, 91:013822, Jan 2015.

\bibitem{LL.V2}
L.D. Landau and E.M. Lifshitz.
\newblock The classical theory of fields.
\newblock {\em Elsevier, Oxford}, 1975.

\bibitem{Tamburini:2010}
M.~Tamburini, F.~Pegoraro, A.~Di~Piazza, C.H. Keitel, and A.~Macchi.
\newblock Radiation reaction effects on radiation pressure acceleration.
\newblock {\em New J. Phys.}, 12(12):123005, 2010.

\bibitem{KravetsRR}
Yevgen Kravets, Adam Noble, and Dino Jaroszynski.
\newblock Radiation reaction effects on the interaction of an electron with an
  intense laser pulse.
\newblock {\em Phys. Rev. E}, 88:011201, Jul 2013.

\bibitem{Ilderton:2013tb}
Anton Ilderton and Greger Torgrimsson.
\newblock {Radiation reaction in strong field QED}.
\newblock {\em Phys. Lett. B}, 725:481, 2013.

\bibitem{gonoskov.phd.2013}
Arkady Gonoskov.
\newblock Ultra-intense laser-plasma interaction for applied and fundamental
  physics.
\newblock {\em Ph.D. thesis, Umea University}, 2013.

\bibitem{bastrakov.jcs.2012}
S.~Bastrakov, R.~Donchenko, A.~Gonoskov, E.~Efimenko, A.~Malyshev, Meyerov. I.,
  and I.~Surmin.
\newblock Particle-in-cell plasma simulation on heterogeneous cluster systems.
\newblock {\em J. Comput. Sci.}, 3(6):474 -- 479, 2012.

\bibitem{capdessus2013influence}
R~Capdessus, E~d’Humieres, and VT~Tikhonchuk.
\newblock Influence of ion mass on laser-energy absorption and synchrotron
  radiation at ultrahigh laser intensities.
\newblock {\em Phys. Rev. Lett.}, 110(21):215003, 2013.

\bibitem{boris1970relativistic}
J.P. Boris.
\newblock Relativistic plasma simulation-optimization of a hybrid code.
\newblock In {\em Proc. Fourth Conf. Num. Sim. Plasmas, Naval Res. Lab, Wash.
  DC}, pages 3--67, 1970.

\end{thebibliography}
\bibliographystyle{unsrt}

\end{document}